\documentclass[12pt]{article}
\usepackage{epsf}
\usepackage{graphicx}
\usepackage{pazh}
\tightenlines

\begin{document}

\title{\bf A hard X-ray survey of the Crux Galactic spiral arm tangent. A
catalog of sources} 

\author{\bf \hspace{-1.3cm}\copyright\, 2005 Ç. \ \
 M.G. Revnivtsev\affilmark{1,2}, S.Yu. Sazonov\affilmark{1,2},
 S.V. Molkov\affilmark{1}, A.A. Lutovinov\affilmark{1},
E.M. Churazov \affilmark{1,2}, R.A. Sunyaev \affilmark{1,2}}
\affil{
$^1$ {\it Space Research Institute, Moscow, Russia}\\
$^2$ {\it Max-Planck-Institut fuer Astrophysik, Garching, Germany }\\
} 

\vspace{2mm}

\sloppypar  
\vspace{2mm}
\noindent
This work is part of a large solid angle hard X-ray survey. We
analized a number of observations by the IBIS telescope aboard the 
INTEGRAL observatory covering the Crux Galactic spiral arm tangent. 
We have detected 46 hard X-ray sources, with 15 of them being new.  Among the
identified sources there are 12 AGNs, 11 HMXBs, 6 LMXBs and 2 active stars. 13
sources remain unidendified.

\noindent

\vfill
 
{$^{*}$ e-mail: revnivtsev@hea.iki.rssi.ru}
\newpage
\thispagestyle{empty}
\setcounter{page}{1}

\section*{Introduction}
The INTEGRAL observatory (Winkler et al. 2003) has devoted a considerable
amount of time to observing Galactic hard X-ray sources. The wide field of
view  ($\sim 15^\circ\times15^\circ$) and large effective area
($\sim2400$ cm$^2$) of the main telescope of the observatory IBIS
permits to survey large areas of the sky resolving individual point
sources. In particular, a record sensitivity was achieved during
surveys of the Galactic Center region (Revnivtsev et al. 2004a), the
Saggitarius spiral arm tangent region (Molkov et al. 2004), and the
Coma cluster region (Krivonos et al. 2005a). A shallower survey of the
Galactic plane was presented by Bird et al. (2004). 

The high penetrating power of hard X-rays seen by the IBIS telescope
permits to study objects previousely hardly detectable by telescopes
operating in soft or standard X-ray energy bands. Thanks to this,
a number of Galactic binary systems with high intrinsic
photoabsorption have been discovered (see e.g. Schartel et al. 2003,
Revnivtsev 2003, Kuulkers 2005, Lutovinov et al. 2005). The optical
companions in such systems are usually quite young and massive and
have powerful stellar winds that form an enshrounding cocoon around
the X-ray source. The INTEGRAL observations have thus strongly
increased the number of known massive binary X-ray systems.

A region of the Galaxy close to the tangent to the Crux spiral 
arm, relatively poorly covered by previous INTEGRAL observations, was
the goal of a deep surveying campaign during INTEGRAL AO-3 cycle
(observations 0320102). The large exposure time has allowed us to 
achieve a sensitivity 0.8--1.0 mCrab in the energy band 17--60
keV. In this paper we present a catalog of detected sources, which can
be subsequently used for source identification and study in different
energy bands. 

\section*{Observations}

We used all publicly available observations by IBIS/INTEGRAL covering
a $50^\circ\times50^\circ$ region of the sky around the position
$l=305, b=0.0$. The main contributions to the net exposure come from
observational set 0320102 (deep survey of the Crux spiral arm
tangent) and from scans of the Galactic plane performed during
2003--2004 as part of the Core program of INTEGRAL. At the edges of
the region a considerable contribution comes from deep observations of
Large Magellanic Cloud and supernova remnant SN1006. The total
exposure time used in our analysis is $\sim$2 million seconds.

The data analysis and detection of sources were performed using
methods described in Revnivtsev et al. (2004a) and Molkov et
al. (2004). The achieved high quality of sky images allows us to lower
the detection limit (the flux at which the probability of a false
source detection is of the order or 100\%) down to $\sim 5\sigma$,
i.e. virtually to the pure statistical limit taking into account
the number of independent pixels on the image. The corresponding
sensitivity contours are presented in Fig.1. 

\section*{Results}

In total 46 sources have been detected. Out of them 41 sources show a
statistically significant flux on the time averaged map. A number of
sources (5 in total) were detected only during particular observations. The
list of detected sources is presented in Table 1. The accuracy of
source localization worsens with decreasing statistical significance
and is $\sim 6\arcmin$ (radius of the 90\% confidence contour) for the
weakest sources.

Search for correlations of detected sources with current X-ray catalogs
(e.g. the X-ray binary catalog, Liu et al. 2000, 2001; ROSAT bright
source catalog, Voges et al. 1999; RXTE catalog, Revnivtsev et
al. 2004b) has allowed us to identify 34 sources. For identification
of IGR J12026-5349 observations by the CHANDRA satellite were used
(Sazonov et al. 2005, in preparation). The accurate localization of a
bright X-ray counterpart found by CHANDRA led to a firm association of
the INTEGRAL source with the active nucleus of the galaxy WKK~0560.

In our sample there 12 active galactic nuclei, 11 Galactic high mass 
X-ray binaries, 6 low mass X-ray binaries, 1 active binary system of
RS CVn type, and one source is a likely symbiotic star. 13 sources 
remain unidintified. 15 new hard X-ray have been discovered.

\bigskip

This work is partly supported by Minpromnauki (grant of President of Russian 
Federation NSH-2083.2003.2), the program of the Russian Academy of Sciences 
``Non stationary phenomena in astronomy'' and RFBR grant 04-02-17276. 
The authors are grateful to the INTEGRAL Science Data Center (Versoix,
Swiss) and the INTEGRAL Russian Science Data Center (Moscow,
Russia). The research is based on observations with INTEGRAL, an ESA
project with instruments and science data center funded by ESA member states
(especially the PI countries: Denmark, France, Germany, Italy,
Switzerland, Spain), Czech Republic and Poland, and with the
participation of Russia and the USA. Some of the results reported in this paper
have been derived using the HEALPix
(http://www.eso.org/science/healpix/, Gorski, Hivon, and Wandelt 
1999) package.

\pagebreak

\pagebreak

{\bf Table 1. }{List of sources detected in the deep survey of the
Galactic plane region near the Crux spiral arm tangent. Sources are
ordered according to the statistical significance of their
detection. References: (1) -- Chernyakova et al. 2005, (2) -- Masetti
et al. 2005, (3) -- Grebenev et al. 2004, (4) -- Produit et al. 2004,
(5) -- Chernyakova et al. 2005, in press; (6) -- Krivonos et
al. 2005b, (7) -- Lubinski et al. 2005, (8) -- Negueruela et al. 2005.
1 mCrab corresponds to an energy flux of $1.4\times10^{11}$
erg/s/cm$^{2}$ for a power law spectrum with a photon index $\Gamma=2.1$} 
\begin{table*} 
\small
{
\begin{tabular}{r|c|c|c|c|c|c}
& Source& R.A.&Dec.&$\sigma$&Flux (17--60 keV), mCrab&Comments (Refs)\\
\hline

 1 &    GX 301-2     & 186.65 & -62.77 & 690.48 & $ 96.9 \pm   0.2$&  HMXB,P 
\\
 2 &  XTE J1550-564  & 237.75 & -56.45 & 405.01 & $ 73.1 \pm   0.2$&  LMXB,BH 
\\
 3 &    Cen A        & 201.36 & -43.01 & 189.12 & $ 37.6 \pm   0.2$&  AGN \\ 
 4 &    Cen X-3      & 170.30 & -60.62 & 172.20 & $ 26.4 \pm   0.2$&  HMXB,P 
\\
 5 & A1145.1-6141    & 176.87 & -61.95 & 131.99 & $ 19.1 \pm   0.2$&  HMXB,P  
\\
 6 & 4U 1538-522     & 235.58 & -52.37 & 111.37 & $ 18.3 \pm   0.2$&  HMXB,P 
\\
 7 & Circinus galaxy & 213.28 & -65.34 & 70.14 & $ 11.9 \pm   0.2$&  AGN \\ 
 8 & NGC 4945        & 196.37 & -49.46 & 61.79 & $ 12.2 \pm   0.2$&  AGN \\ 
 9 & 4U 1516-569     & 230.17 & -57.17 & 61.02 & $ 11.4 \pm   0.2$&  LMXB,B \\ 
10 & PSR 1509-58     & 228.47 & -59.14 & 43.67 & $  8.6 \pm   0.2$&  PSR \\ 
11 & 4U 1323-619     & 201.64 & -62.13 & 37.93 & $  5.4 \pm   0.2$&  LMXB \\ 
12 & 4U 1626-67      & 248.06 & -67.47 & 27.07 & $ 12.5 \pm   0.5$&  LMXB,P \\ 
13 & NGC 4507        & 188.90 & -39.91 & 25.45 & $  8.3 \pm   0.4$&  AGN \\ 
14 & 4U 1344-60      & 206.89 & -60.61 & 23.50 & $  3.6 \pm   0.2$&  AGN \\ 
15 & IGR J12346-6434 & 188.71 & -64.56 & 22.34 & $  3.2 \pm   0.2$&  
Symb.st.(1,2)\\
16 & 4U 1246-588     & 192.39 & -59.08 & 22.35 & $  3.2 \pm   0.2$&  XB \\ 
17 & X1145-619       & 176.99 & -62.20 & 20.82 & $  3.1 \pm   0.2$&  HMXB,P \\ 
18 & IGR J11435-6109 & 175.99 & -61.12 & 18.68 & $  2.7 \pm   0.2$&  HMXB,P, 
(3) \\
19 & IGR J11305-6256 & 172.77 & -62.94 & 16.26 & $  2.5 \pm   0.2$&  - (4)\\ 
20 & 2S 1254-690     & 194.41 & -69.29 & 12.18 & $  2.3 \pm   0.2$&  LMXB \\ 
21 & 1RXP J130159.6-635806 & 195.50 & -63.96 & 11.74 & $  1.7 \pm   0.2$&  
HMXB,P (5)\\
22 & IGR J12026-5349 & 180.67 & -53.83 & 11.72 & $  2.1 \pm   0.2$&
 AGN,WKK 0560 \\ 
23 & 4U 1543-624     & 236.94 & -62.55 & 11.59 & $  2.7 \pm   0.3$&  LMXB,NS 
\\
24 & IGR J14579-4308 & 224.44 & -43.14 & 11.11 & $  1.8 \pm   0.2$&  AGN:IC 
4518 \\
25 & XTE J1543-568   & 236.01 & -56.75 &  9.97 & $  1.8 \pm   0.2$&  HMXB,P \\ 
26 & NGC 6300        & 259.20 & -62.79 &  9.61 & $  3.1 \pm   0.4$&  AGN \\ 
27 & ESO 323-G077    & 196.60 & -40.42 &  7.58 & $  1.7 \pm   0.2$&  AGN \\ 
28 & XSS J12270-4859 & 186.99 & -48.90 &  7.60 & $  1.8 \pm   0.3$&  -\\ 
29 & IGR J13109-5552 & 197.69 & -55.86 &  7.14 & $  1.1 \pm   0.2$&  -\\ 
30 & IGR J15360-5750 & 234.00 & -57.84 &  6.33 & $  1.2 \pm   0.2$&  -\\ 
31 & IGR J14175-4641 & 214.30 & -46.68 &  6.11 & $  1.2 \pm   0.2$&  -\\ 
32 & IGR J16185-5928 & 244.63 & -59.47 &  5.70 & $  1.2 \pm   0.2$&  AGN:WKK 
6471 \\
33 & IGR J10109-5746 & 152.71 & -57.78 &  5.57 & $  1.3 \pm   0.3$&  - \\ 
34 & IGR J14493-5534 & 222.29 & -55.60 &  5.56 & $  1.0 \pm   0.2$&  - \\ 
35 & IGR J10252-6829 & 156.30 & -68.46 &  5.42 & $  1.6 \pm   0.3$&  - \\ 
36 & IGR J14552-5133 & 223.81 & -51.55 &  5.37 & $  1.0 \pm   0.2$&
 AGN:WKK 4438 \\ 
37 & 1ES 1210-646    & 183.24 & -64.84 &  5.15 & $  0.8 \pm   0.2$&  - \\ 
38 & IGR J08023-6954 & 120.57 & -69.91 &  5.14 & $  1.7 \pm   0.4$&  - \\ 
39 & IGR J14471-6319 & 221.78 & -63.33 &  5.09 & $  0.9 \pm   0.2$&  -  \\
40 & IGR J15094-6649 & 227.32 & -66.85 &  5.02 & $  1.1 \pm   0.2$&  -  \\
41$^*$ & IGR J11321-5311 & 173.03 & -53.18 &  10.5 & $  27  \pm   2.6$&  -(6)\\
42$^*$ & IGR J11215-5952 & 170.42 & -59.90 &  8.42 & $  1.8 \pm   0.2$&  HMXB 
(7,8)\\
43$^*$ & HD 101379       & 174.88 & -65.33 &  7.07 & $  1.6 \pm   0.2$&  RS 
CVn \\
44$^*$ & PSR B1259-63    & 195.70 & -63.83 &  6.36 & $  0.9 \pm   0.2$&  
HMXB,P \\
45$^*$ & IGR J11085-5100 & 167.14 & -51.01 &  5.64 & $  1.9 \pm   0.4$&  - \\ 
46$^*$ & 4U 1022-55      & 159.45 & -56.74 &  5.54 & $  1.3 \pm   0.3$&  HMXB 
\\
47$^*$ & IGR J12415-5750 & 190.35 & -57.84 &  5.00 & $  1.0 \pm   0.2$&  
AGN:WKK 1263\\
\hline
\end{tabular}
}
\begin{list}{}
\item $^{*}$ -- these sources are not statistically significantly detected
on the time averaged map, but are detected in subsets of observations.
In column 5 the maximum detection significance is quoted.
\end{list}
\end{table*}
\newpage

\begin{figure}
\includegraphics[width=\textwidth,angle=-90]{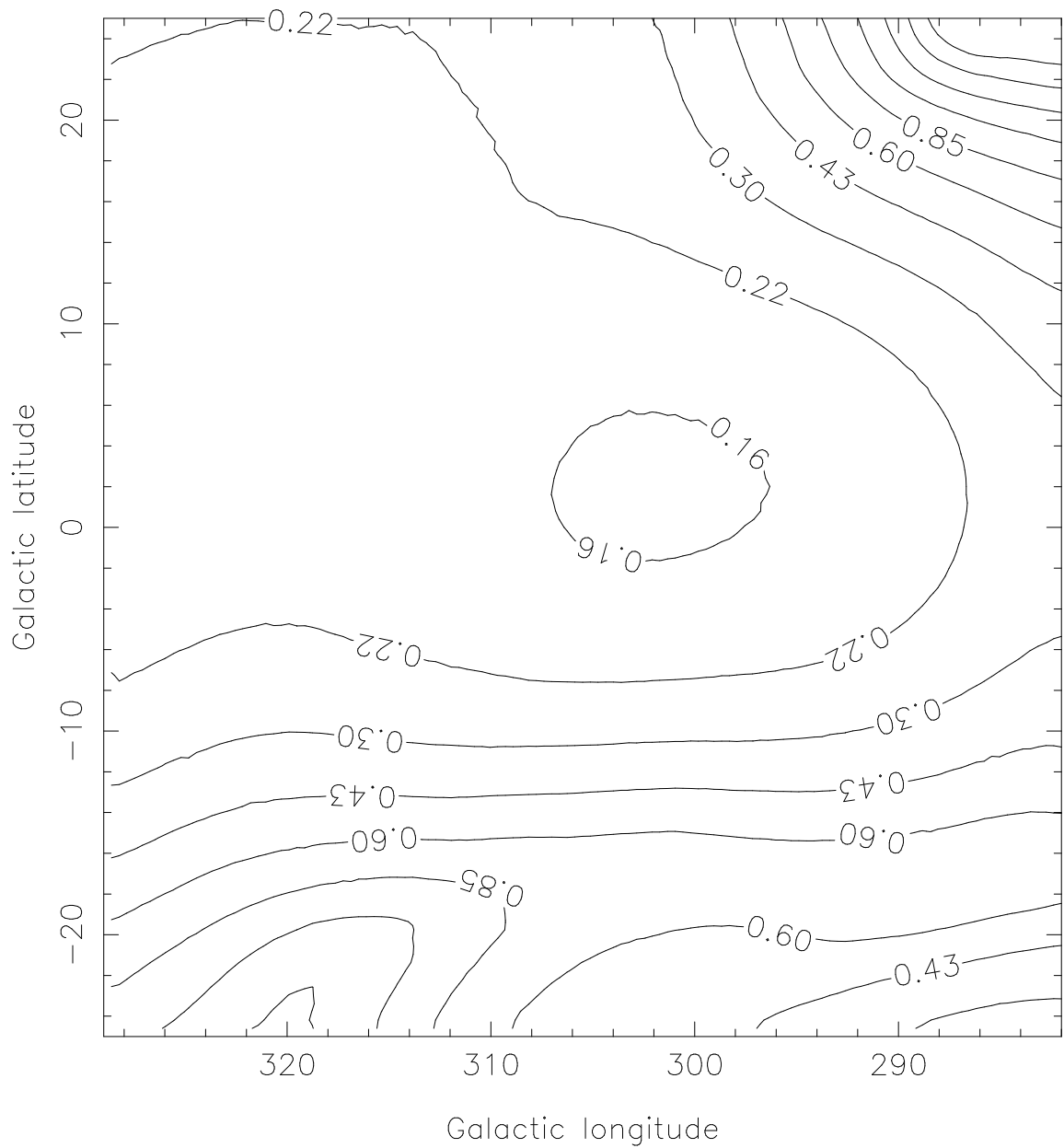}
\caption{Statistical uncertainty in flux measurement over the sky
region covered by the survey. Contours show regions where 1$\sigma$
statistical error is 0.16, 0.22, 0.3, etc. mCrab} 
\end{figure}

\begin{figure}
\includegraphics[width=\textwidth]{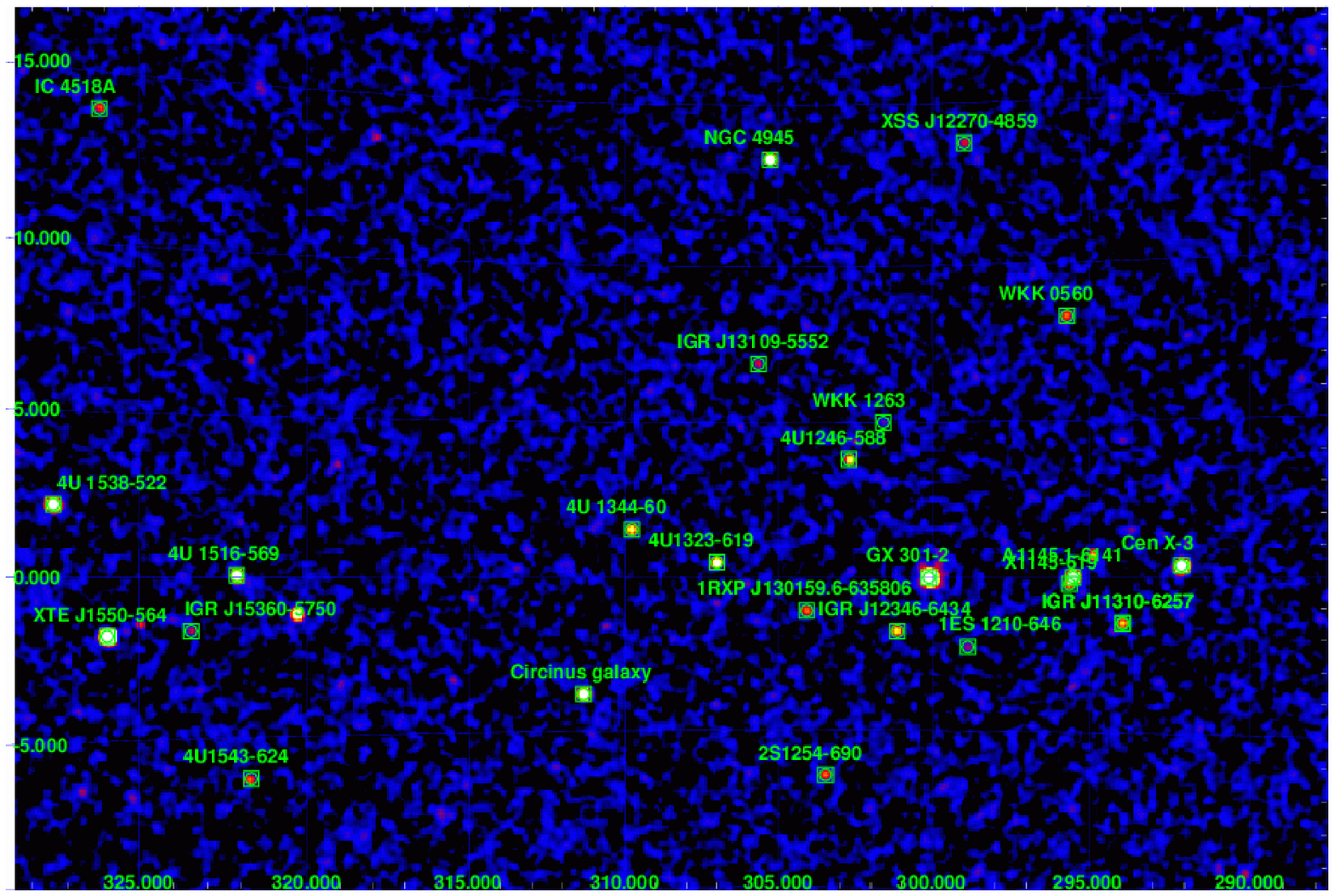} 
\caption{Map of the galactic plane near the tangent to the Crux 
spiral arm in the energy band 17--60 keV. Only brightest sources are marked}
\end{figure}


\begin{references}

A. Bird, E. Barlow, L. Bassani et al., Astroph. J., {\bf 607}, 33L (2004)

M. Chernyakova, T.J.-L. Courvoisier, J. Rodriguez, A. Lutovinov,
Astronomer's Telegram 519 (2005) 

S. Grebenev, P. Ubertini, J. Chenevez et al., Astronomer's Telegramm 350 (2004)

R. Krivonos, A. Vikhlinin, E. Churazov, A. Lutovinov, S. Molkov,
R. Sunyaev, Astroph. J. {\bf 625}, 89 (2005)

E. Kuulkers, in Proc. of the Interacting Binaries Meeting of Cefalu,
Italy, July 2004 (astro-ph/0504625)

Q.Z. Liu, J. van Paradijs, E.P.J. van den Heuvel,
Astr. Astroph. Suppl., {\bf 147}, 25 (2000) 

Q.Z. Liu, J. van Paradijs, E.P.J. van den Heuvel, Astr. Astroph., {\bf
368}, 1021 (2001)

A. Lutovinov,  M. Revnivtsev, M. Gilfanov, P. Shtykovskiy, S. Molkov, 
R. Sunyaev, Astron. Astroph, in press (2005), astro-ph/0411550 

P. Lubinski, M. Gadolle Bel, A. von Kienlin et al., Astronomer's
Telegram 469 (2005) 

N. Masetti, L. Bassani, A.J. Bird, A. Bazzano,  Astronomer's Telegram
528 (2005) 

S.V. Molkov, A.M. Cherepashchuk, A.A. Lutovinov, M.G. Revnivtsev,
K.A. Postnov, R.A. Sunyaev, Astron. Lett, {\bf 30}, 534 (2004) 

I. Negueruela, D.M. Smith, S. Chaty, Astronomer's Telegram 470 (2005)

N. Produit, J. Ballet, N. Mowlavi, Astronomer's Telegram 278 (2005)

M.G. Revnivtsev, Astron. Lett. {\bf 29}, 644 (2003)

M.G. Revnivtsev, R.A. Sunyaev, D.A. Varshalovich et al.,
Astron. Lett.,{\bf 30}, 382 (2004a) 

M.G. Revnivtsev, S. Sazonov, K. Jahoda, M. Gilfanov, Astron. Astroph.,
{\bf 418}, 927 (2004b) 

N. Schartel, M. Ehle, M. Breitfellner et al., IAU Circ. 8072 (2003).

W. Voges, B. Aschenbach, Th. Boller et al., Astron. Astroph., {\bf
349}, 389 (1999) 

C. Winkler, T.J.-L. Courvoisier, G. Di Cocco et
al. Astron. Astroph. {\bf 411}, L1 (2003)

\end{references}
\end{document}